\documentstyle[aps,eqsecnum,preprint,floats,epsf,psfig]{revtex}
\begin{document}

\def\be{\begin{eqnarray}}
\def\en{\end{eqnarray}}
\def\up{\uparrow}
\def\dw{\downarrow}
\def\non{\nonumber}
\def\la{\langle}
\def\ra{\rangle}
\def\etapp{{\eta^{(')}}}
\def\ov{\overline}
\def\vp{\varepsilon}
\def\half{{{1\over 2}}}
\def\a{{\cal A}}
\def\b{{\cal B}}
\def\c{{\cal C}}
\def\d{{\cal D}}
\def\pr{{\sl Phys. Rev.}~}
\def\prl{{\sl Phys. Rev. Lett.}~}
\def\pl{{\sl Phys. Lett.}~}
\def\np{{\sl Nucl. Phys.}~}
\def\zp{{\sl Z. Phys.}~}

\font\el=cmbx10 scaled \magstep2
{\obeylines
\hfill }

\vskip 1.5 cm

\centerline{\large\bf Exclusive Hadronic $D$ Decays to $\eta'$ or
$\eta$:} \centerline{\large\bf Addendum on Resonant Final-State
Interactions}
\medskip
\bigskip
\medskip
\centerline{\bf Hai-Yang Cheng and B. Tseng}
\medskip
\centerline{Institute of Physics, Academia Sinica}
\centerline{Taipei, Taiwan 115, Republic of China}
\medskip

\bigskip
\bigskip
\bigskip
\centerline{\bf Abstract}
\bigskip
{\small For hadronic two-body decays of charmed mesons involving
$\eta$ or $\eta'$, resonance-induced final-state interactions
(FSI) that mimic the $W$-exchange or the $W$-annihilation topology
can play an essential role. In particular, the decays $D^0\to\ov
K^0\eta'$ and $D^+\to\pi^+\eta$, which are largely suppressed in
the absence of FSI, are enhanced dramatically by resonant FSI. It
is stressed that the effect of resonant FSI is negligible for
$\rho^+(\eta,\eta')$ final states because of the mismatch of the
$G$ parity of $\rho^+(\eta,\eta')$ and the $J=0$, $I=1$ meson
resonance. We argue that a possible  gluon-mediated process in
which two gluons couple directly to the gluonic component of the
$\eta'$, e.g. the gluonium, rather than to the flavor-singlet
$\eta_0$, can enhance both modes $D_s^+\to\rho^+\eta'$ and
$D_s^+\to\rho^+\eta$, especially the former; that is, this new
mechanism can account for the unexpectedly large branching ratio
of $\rho^+\eta'$ without suppressing $\rho^+\eta$.

}

\pagebreak

\section{Introduction}
We have shown recently in \cite{CT99} that in the decays of
charmed mesons into the final states containing an $\eta$ or
$\eta'$, final-state interactions (FSI) in the resonance formation
are able to enhance $\b(D^0\to\ov K^0\eta')$ and
$\b(D^+\to\pi^+\eta)$ by an order of magnitude. Resonance-induced
couple-channel effects will suppress $D_s^+\to\pi^+\eta$ and
enhance $D_s^+\to\pi^+\eta'$. Contrary to $D\to P\etapp$ decays,
resonant FSI play only a minor role for $D^0\to\ov K^{*0}\etapp$
and do not contribute to $(D^+,D_s^+)\to\rho^+ \etapp$. We argued
that it is difficult to understand the observed large decay rates
of the $\rho^+\eta'$ and $\rho^+\eta$ decay modes of $D_s^+$
simultaneously. FSI are not helpful due to the absence of
$D_s^+\to PP$ decays that have much larger decay rates than
$D_s^+\to\rho^+\eta'$. $W$-annihilation and a possible production
of the $\eta'$ due to gluon-mediated processes can in principle
enhance $\b(D_s^+\to\rho^+\eta')$, but, unfortunately, they will
also suppress $\b(D_s^+\to\rho^+\eta)$.

In \cite{CT99} we have followed \cite{Zen} to use the strong
reaction matrix $K_0$ together with the unitarity constraint of
the $S$ matrix to study the effects of resonant FSI and showed
that resonance-induced FSI amount to modifying, for example, the
$W$-exchange amplitude $\c$ in $D^0\to \ov K\pi,\ov K\eta,\ov
K\eta'$ decays by \cite{Zen} \be \label{newc} \c\to \c+(\c+{1\over
3}\a)(\cos\delta e^{i\delta}-1)
\en
and leaving the other quark-diagram amplitudes intact, where $\a$
is an external $W$-emission amplitude. In this addendum we will
derive the above relation in a rigorous way and find that the
modification due to FSI for the $W$-exchange amplitude as shown in
Eq. (\ref{newc}) is too small by a factor of 2.

In the present paper we will update the previous analysis
\cite{CT99} by correcting the error occurred in Eq. (\ref{newc}),
discussing its implication and employing the new measurement of
the $D_s^+$ lifetime. Moreover, we shall show explicitly that
contributions from resonant FSI to the decays
$(D^+,D_s^+)\to\rho^+\etapp$ should be negligible, otherwise the
predicted branching ratios will become too large compared to
experiment. This is ascribed to the mismatch of the $G$ parity of
$\rho^+(\eta,\eta')$ and the $J=0$, $I=1$ meson resonance. We will
also employ the mode $D^0\to\ov K^0\pi^0$ as an example to
demonstrate that resonance-induced FSI, which are crucial for some
two-body decays involving one single isospin component, e.g. the
final state containing an $\eta$ and $\eta'$, play only a minor
role compared to isospin FSI for decay modes involving several
different isospin components. We then turn to some possible
explanation of the unexpectedly large branching ratio of
$D_s^+\to\rho^+\eta'$. Finally we discuss in detail the possible
sources of theoretical uncertainties for estimating the effects of
resonant FSI.

\section{Resonant Final-State Interactions}
There are several different forms of FSI: elastic scattering and
inelastic scattering such as quark exchange, resonance
formation,$\cdots$, etc. Since FSI are nonperturbative in nature,
in general it is notoriously difficult to calculate their effects.
Nevertheless, the effect of resonance-induced coupled-channel FSI
can be estimated provided that the mass and the width of the
nearby resonances are known. It appears that the resonance
formation of FSI via $q\bar q$ resonances is probably the most
important one if the final state has only one single isospin
component. For previous studies of the effects of resonant FSI in
charm decays, see \cite{Buccella,Zen,CT99,Fajfer,aaoud}.

In the presence of resonances, the decay amplitude of the charmed
meson $D$ decaying into two mesons $M_1M_2$ is modified by
rescattering through a multiplet of resonances
\cite{Weinberg}\footnote{The same expression is also given in
\cite{Buccella} except that the phase in Eq. (3.3) of
\cite{Buccella} is too small by a factor of 2.}
\be
\label{FSI}A(D\to M_iM_j)^{\rm FSI}=A(D\to M_iM_j)-i{\Gamma\over
E-m_R+i\Gamma/2}\sum_r c^{(r)}_{ij}\sum_{kl} c^{(r)*}_{kl}A(D\to
M_kM_l),
\en
where $c^{(r)}_{ij}$ are normalized coupling constants of $M_iM_j$
with the scalar resonance $r$, obeying the relations
\be
\sum_{ij}c_{ij}^{(r)}c_{ij}^{(s)*}=\delta_{rs}, \qquad
\sum_{ij}|c_{ij}^{(r)}|^2=1.
\en
The presence of a resonance shows itself in a characteristic
behavior of phase shifts near the resonance. For each individual
resonant state $r$, there is an eigenstate of $A(D\to M_iM_j)$
with eigenvalue \cite{Weinberg}
\be \label{phase}
e^{2i\delta_r}=1-i\,{\Gamma\over m_D-m_R+i\Gamma/2},
\en
in the rest frame of the charmed meson, where $m_R$ and $\Gamma$
are the mass and the width of the resonance, respectively.
Therefore, resonance-induced FSI are amenable technically in terms
of the physical resonances.

\subsection{$D^0\to (\ov K^0,\ov K^{*0})(\eta,\eta')$ decays}
To illustrate the effect of FSI in the resonance formation,
consider the decays $D^0\to \ov K_i P_j$ as an example. The only
nearby $0^+$ scalar resonance with $(s\bar d)$ quark content in
the charm mass region is $r=K_0^*(1950)$ and the states $\ov K_i
P_j$ are $K^-\pi^+,\ov K^0\pi^0,\ov K^0\eta,\ov K^0\eta'$. The
quark-diagram amplitudes for $D^0\to K^-\pi^+,~\ov K^0\pi^0,~\ov
K^0\eta_{ns}$ and $\ov K^0\eta_s$, where $\eta_{ns}={1\over
\sqrt{2}}(u\bar u+d\bar d)$ and $\eta_s=s\bar s$, are given by
(see Table III of \cite{CC87}): \be \label{qds} && A(D^0\to(\ov
K\pi)_{3/2}) = {1\over\sqrt{3}}(\a+\b),   \qquad A(D^0\to(\ov
K\pi)_{1/2}) = {1\over\sqrt{6}}(2\a-\b+3\c),   \non
\\ && A(D^0\to\ov K^0\eta_{ns}) = {1\over\sqrt{2}}(\b+\c), \qquad
A(D^0\to\ov K^0\eta_s) = \c,
\en
where the subscripts 1/2 and 3/2 denote the isospin of the $\ov K
\pi$ system. In Eq. (\ref{qds}), $\a$ is the external $W$-emission
amplitude, $\b$ the internal $W$-emission amplitude and $\c$ the
$W$-exchange amplitude \cite{CC87}.\footnote{The quark-diagram
amplitudes $\a,\,\b,\,\c,\,\d$ for external $W$-emission, internal
$W$-emission, $W$-exchange and $W$-annihilation are sometimes
denoted by $T,\,C,\,E,\,A$, respectively, in the literature.}
Consider the $D$-type coupling for the strong interaction
$P_1P_2\to P'$ ($P'$: scalar meson), namely $\kappa{\rm
Tr}\left(P'\{P_1,\,P_2\}\right)$ with $\kappa$ being a
flavor-symmetric strong coupling \cite{Zen}. Noting that $(\ov
K\pi)_{3/2}$ does not couple to $(\ov K\pi)_{1/2}$, $\ov
K^0\eta_{ns}$, and $\ov K^0\eta_s$ via FSI, the matrix $c^2$
arising from two $D$-type couplings in the $I={1\over 2}$ sector
has the form:
\be
c^2 \propto \kappa^2\left(\matrix{ {3\over 2} & {\sqrt{3}\over 2}
& {\sqrt{3}\over\sqrt{2}} \cr   {\sqrt{3}\over 2} & {1\over 2} &
{1\over \sqrt{2}} \cr    {\sqrt{3}\over\sqrt{2}}  &
{1\over\sqrt{2}} & 1    \cr} \right)
\en
in the basis of $(\ov K\pi)_{1/2},~\ov K^0\eta_{ns},~\ov
K^0\eta_s$. Hence, the normalized matrix $c^2$ is given by
\be
c^2=\left(\matrix{ {1\over 2} & {1\over 2\sqrt{3}} & {1\over
\sqrt{6}}  \cr {1\over 2\sqrt{3}} & {1\over 6} & {1\over
3\sqrt{2}} \cr {1\over\sqrt{6}} & {1\over 3\sqrt{2}} & {1\over 3}
\cr}\right).
\en
Then it is easily seen that
\be
A(D^0 &\to& \ov K^0\eta_s)^{\rm FSI} = \c^{\rm
FSI}=\c+(e^{2i\delta_r}-1) \cr & \times &
\left[{1\over\sqrt{6}}A(D^0\to (\ov K\pi)_{1/2})+{1\over
3\sqrt{2}}A(D^0\to \ov K^0\eta_{ns})+{1\over 3}A(D^0\to\ov
K^0\eta_s)\right],
\en
and hence
\be
\c^{\rm FSI}=\c+(e^{2i\delta_r}-1)\left(\c+{\a\over 3}\right).
\en
Therefore, resonance-induced FSI amount to modifying the
$W$-exchange amplitude and leaving the other quark-diagram
amplitudes $\a$ and $\b$ intact. Since
$(e^{2i\delta_r}-1)=2(\cos\delta_r e^{i\delta_r}-1)$, we see that
the contribution of resonant FSI to the $W$-exchange amplitude as
given in Eq. (\ref{newc}) is too small by a factor of 2.

The resonance contribution to FSI, which arises mainly from the
external $W$-emission diagram for the decay $D^0\to (\ov
K\pi)_{1/2}$ followed by final-state $q\bar q$ resonance, has the
same topology as the $W$-exchange quark diagram. We thus see that
even if the short-distance $W$-exchange vanishes, as commonly
asserted, an effective long-distance $W$-exchange still can be
induced via FSI in resonance formation.

Considering the $\eta-\eta'$ mixing parameterized by
\be
\eta'=\eta_8\sin\theta+\eta_0\cos\theta, \qquad
\eta=\eta_8\cos\theta-\eta_0 \sin\theta,
\en
with $\eta_8$ and $\eta_0$ being SU(3) octet and singlet wave
functions respectively, and neglecting the $W$-exchange amplitude
$\c$, we obtain \cite{CT99} \be \label{nDKeta}
 A(D^0\to\ov K^0\eta) = {G_F\over\sqrt{2}}V_{cs}^*
V_{ud}\left[a_2X^{(D^0\eta,\ov K^0)}+ a_1X^{(D^0 K^-,\pi^+)}\,{
e^{2i\delta_r}-1\over 3\sqrt{6}}\left(
-\cos\theta-2\sqrt{2}\sin\theta\right)\right], \non\\
 A(D^0\to\ov K^0\eta') = {G_F\over\sqrt{2}}V_{cs}^*
V_{ud}\left[a_2X^{(D^0\eta',\ov K^0)}+ a_1X^{(D^0K^-,\pi^+)}\,{
e^{2i\delta_r}-1\over 3\sqrt{6}}\left(
-\sin\theta+2\sqrt{2}\cos\theta\right)\right], \non\\
\en
and
\be \label{nDK*eta} A(D^0\to \ov K^{*0}\eta) &=&
{G_F\over\sqrt{2}}V_{cs}^*V_{ud}\Bigg[ a_2X^{( D^0\eta,\ov
K^{0*})} \non
\\ &+& a_1\left(X^{(D^0K^{-*},\pi^+)}+X^{(D^0K^-,\rho^+)}\right)\,{
e^{2i\delta_{r'}}-1\over 6\sqrt{6}}\left(
-\cos\theta-2\sqrt{2}\sin\theta\right)\Bigg], \non\\ A(D^0\to \ov
K^{*0}\eta') &=& {G_F\over\sqrt{2}}V_{cs}^*V_{ud}\Bigg[ a_2X^{(
D^0\eta',\ov K^{0*})}   \non
\\ &+& a_1\left(X^{(D^0K^{-*},\pi^+)}+X^{(D^0K^-,\rho^+)}\right)\,{
e^{2i\delta_{r'}}-1\over 6\sqrt{6}}\left(
-\sin\theta+2\sqrt{2}\cos\theta\right)\Bigg],
\en
where $X^{(DM_1,M_2)}$ denotes the factorizable amplitude with the
meson $M_2$ being emitted out:
\be
X^{(DM_1,M_2)}=\la M_2|(\bar q_1 q_2)|0\ra\la M_1|(\bar q_3
c)|D\ra,
\en
with $(\bar q_1q_2)\equiv\bar q_1\gamma_\mu(1-\gamma_5)q_2$.
Explicitly,
\be
X^{(D^0\etapp,\ov K^0)} &=&
if_K(m_D^2-m^2_\etapp)F_0^{D^0\etapp}(m^2_K), \non \\ X^{(D^0K^-
,\pi^+)} &=& if_K(m_D^2-m^2_K)F_0^{D^0K^-}(m^2_\pi), \non \\
X^{(D\etapp,K^*)} &=&
2f_{K^*}m_{K^*}F_1^{D\etapp}(m^2_{K^*})(\vp\cdot p_{_D}), \non \\
X^{(D^0K^-,\rho^+)} &=&
2f_{\rho}m_{\rho}F_1^{D^0K^-}(m^2_{\rho})(\vp\cdot p_{_D}), \non
\\
X^{(D^0K^{*-},\pi^+)} &=& 2f_\pi m_{K^*} A_0^{D^0 K^{-*}}(m^2_\pi)
(\vp\cdot p_{_D}),
\en
where the form factors $F_0,~F_1$ and $A_0$ are those defined in
\cite{BSW}.

Since $F_0^{D\eta'}(0)<F_0^{D\eta}(0)$ \cite{CT99} and the
available phase space for $K\eta'$ is less than that for $K\eta$,
the factorization approach implies less $\eta'$ production than
$\eta$ in $D^0\to\ov K^0\etapp$ decays, in disagreement with
experiment (see Table I). To see how the mechanism of resonant FSI
works, notice that the big parentheses in Eqs.
(\ref{nDKeta},\ref{nDK*eta}) reflect the coefficient of the
$W$-exchange amplitude. Taking $\theta=-19.5^\circ$ as a
benchmark, the wave functions of the $\eta$ and $\eta'$ have the
simple expressions \cite{Chau91}:
\be
\eta={1\over\sqrt{3}}(u\bar u+d\bar d-s\bar s), \qquad
\eta'={1\over\sqrt{6}}(u\bar u+d\bar d+2s\bar s).
\en
In Eqs. (\ref{nDKeta},\ref{nDK*eta}),
$(\cos\theta+2\sqrt{2}\sin\theta)=0$ indicates that there is no
intrinsic $W$-exchange diagram in $D^0\to \ov K^{0(*)}\eta$, while
$(2\sqrt{2}\cos\theta-\sin\theta)=3$ shows that $D^0\to
K^{0(*)}\eta'$ contains the amplitude $3\,\c$. Since the internal
$W$-emission amplitude is color suppressed, while the
contributions from resonant FSI are induced from the external
$W$-emission, it is clear that the decay $D^0\to \ov K^{0}\eta'$
receives large contributions from FSI in the resonance form, so
that its decay rate is larger than that of $D^0\to\ov K^0\eta$.

Using the effective coefficients
\be \label{a1a2} a_1=1.25, \qquad
a_2=-0.51,
\en
the $\eta-\eta'$ mixing angle $\theta=-22^\circ$ \cite{Holstein},
the mass $1945\pm 10\pm 20$ MeV and the width $210\pm 34\pm 79$
MeV for the $0^+$ resonance $K_0^*(1950)$, $m_R=1830$ MeV and
$\Gamma=250$ MeV for the $0^-$ resonance $K(1830)$ \cite{PDG}, and
various form factors given in \cite{CT99}, the calculated
branching ratios are exhibited in Table I. We see that in the
presence of resonant FSI, the branching ratio of $\ov K^{0*}\eta'$
is enhanced by an order of magnitude, while the decay rate of $\ov
K^{0*}\eta$ is only slightly increased. The $\eta'$ enhancement
for $D^0\to\ov K^0\eta'$ over $D^0\to\ov K^0\eta$, which cannot be
accounted for in the factorization approach, can be explained in
terms of resonance-induced FSI. For comparison, the theoretical
predictions by Buccella {\it et al.,} \cite{Buccella} are also
shown in Table I. As noted in passing, the phase in Eq. (3.3) of
\cite{Buccella} is too small by a factor of 2.

It should be stressed that although resonant FSI can make a
dramatic effect on hadronic decays of the charmed mesons
containing an $\eta$ or $\eta'$, i.e. final states with one single
isospin component, they are not expected to play the same
essential role in the decay channels involving several different
isospin components. A well known example is the decay $D^0\to\ov
K^0\pi^0$ with the decay amplitude:
\be
A(D^0\to \ov K^0\pi^0) &=& a_2X^{(D^0\pi^0,\ov K^0)}-{1\over 3}
a_1(e^{2i\delta_r}-1)X^{(D^0K^-,\pi^+)}  \non \\ &+&
\left(a_1X^{(D^0K^-,\pi^+)}+a_2X^{(D^0\pi^0,\ov
K^0)}\right){\sqrt{2}\over
3}\left(e^{-i(\delta_{1/2}-\delta_{3/2})}-1\right),
\en
where $X^{(D^0\pi^0,\ov
K^0)}=if_K(m_D^2-m_\pi^2)F_0^{D^0\pi^+}(m_K^2)/\sqrt{2}$ and
$\delta_i$ are the isospin phase shifts. In naive factorization
with $a_{1,2}=c_{1,2}+c_{2,1}/3$ and in the absence of any FSI, we
find ${\cal B}(D^0\to\ov K^0\pi^0)=0.03\%$ for $c_1(m_c)=1.26$ and
$c_2(m_c)=-0.51$, which is obviously too small compared to the
experimental value $(2.12\pm 0.21)\%$ \cite{PDG}. In the
large-$N_c$ limit where $a_2=c_2$, the branching ratio is
increased to 1.0\%. When the resonant FSI are turned on, ${\cal
B}(D^0\to\ov K^0\pi^0)$ is {\it decreased} to 0.36\% ! Using the
isospin phase shift difference
$(\delta_{1/2}-\delta_{3/2})=71.4^\circ$ extracted from the
isospin analysis of $D\to\ov K\pi$ data, the branching ratio of
$D^0\to\ov K^0\pi^0$ is enhanced by isospin FSI to 1.44\%. It is
clear that in order to understand the color non-suppression of
$D^0\to\ov K^0\pi^0$, one needs nonfactorizable effects to account
for the non-smallness of $a_2$ and isospin FSI to generate
adequate $\ov K^0\pi^0$ from $K^-\pi^+$.

\subsection{$D^+\to (\pi^+,\rho^+)(\eta,\eta')$ decays}
Proceeding as before, resonance-induced coupled-channel effects
among the three channels: $K^+\ov K^0,\,\pi^+\eta_{ns}$ and
$\pi^+\eta_s$ will only modify the magnitude and phase of the
$W$-annihilation amplitude $\d$ and leave the other quark-diagram
amplitudes unaffected \cite{Zen}:
\be
\d\to \d+\left(\d+{1\over 3}\a\right)(e^{2i\delta_r}-1).
\en
The decay amplitudes of the Cabibbo-suppressed decays $D^+\to
\pi^+\eta^{(')}$ and $\rho^+\eta^{(')}$ in the presence of FSI via
$\bar qq$ resonance are \cite{CT99}:
\be
A(D^+\to\pi^+\eta) &=& {G_F\over\sqrt{2}}V_{cd}^*V_{ud}\Bigg[
a_1X^{(D^+\eta,
\pi^+)}+a_2\left(X_d^{(D^+\pi^+,\eta)}-X_s^{(D^+\pi^+,\eta)}\right)
\non
\\ &+& {1\over 3\sqrt{3}}a_1X^{(D^+\ov K^0,K^+)}(
e^{2i\delta_r}-1)\left(-\sqrt{2}\cos\theta+2\sin\theta
\right)\Bigg], \non \\ A(D^+\to\pi^+\eta') &=&
{G_F\over\sqrt{2}}V_{cd}^*V_{ud}\Bigg[ a_1X^{(D^+\eta',
\pi^+)}+a_2\left(X_d^{(D^+\pi^+,\eta')}-X_s^{(D^+\pi^+,\eta')}\right)
\non
\\ &+& {1\over 3\sqrt{3}}a_1X^{(D^+\ov K^0,K^+)}(
e^{2i\delta_r}-1)\left(-\sqrt{2}\sin\theta-2\cos\theta
\right)\Bigg],
\en
and
\be
A(D^+\to\rho^+\eta) &=& {G_F\over\sqrt{2}}V_{cd}^*V_{ud}\Bigg[
a_1X^{(D^+\eta,
\rho^+)}+a_2\left(X_d^{(D^+\rho^+,\eta)}-X_s^{(D^+\rho^+,\eta)}\right)\Bigg],
\non \\
 A(D^+\to\rho^+\eta') &=&
{G_F\over\sqrt{2}}V_{cd}^*V_{ud}\Bigg[ a_1X^{(D^+\eta',
\rho^+)}+a_2\left(X_d^{(D^+\rho^+,\eta')}-X_s^{(D^+\rho^+,\eta')}\right)
\Bigg],
\en
where
\be
X_q^{(D^+\pi^+,\etapp)} &=&
if^q_\etapp(m_D^2-m^2_\pi)F_0^{D^+\pi^+}(m^2_\etapp), \non \\
X_q^{(D^+\rho^+,\etapp)} &=& 2f^q_\etapp m_\rho
A_0^{D^+\rho^+}(m^2_\etapp) (\vp\cdot p_{_D}),
\en
and the values of the decay constants $f_{\eta^{(')}}^q$ can be
found in \cite{CT99}.

Note that since $\pi^+\pi^0$ does not couple to $\pi^+\eta^{(')}$
by strong interactions, $D^+\to\pi^+\eta^{(')}$ receive
contributions from resonant FSI only through the process $D^+\to
K^+\ov K^0 \to\pi^+\eta^{(')}$. As for the decay
$D^+\to\rho^+\eta^{(')}$ one may naively expect that
\be
A(D^+\to\rho^+\eta) &=& {G_F\over\sqrt{2}}V_{cd}^*V_{ud}\Bigg[
a_1X^{(D^+\eta,
\rho^+)}+a_2\left(X_d^{(D^+\rho^+,\eta)}-X_s^{(D^+\rho^+,\eta)}\right)
\non \\
 &+& {1\over 6\sqrt{3}}a_1\Big(X^{(D^+\ov K^{0*},K^+)}
 + X^{(D^+\ov K^0,K^{+*})}
-\sqrt{2}X^{(D^+\pi^0,\rho^+)}-\sqrt{2}X^{(D^+\rho^0,\pi^+)}\Big)
\non \\ &\times& (
e^{2i\delta}-1)\left(-\sqrt{2}\cos\theta+2\sin\theta
\right)\Bigg],
\en
and likewise for the $\rho^+\eta'$ state, where use of
$V_{cs}^*V_{us}\approx -V^*_{cd}V_{ud}$ has been made. Since the
factorized term $X^{(D^+\pi^0,\rho^+)}\propto \la\pi^0|(\bar
dc)|D^+\ra$ has a sign opposite to that of $X^{(D^+\ov
K^{0},K^{+*})}$ due to the pion wave function $\pi^0=(u\bar
u-d\bar d)/\sqrt{2}$, there is no cancellation among various
contributions to resonant FSI. Employing $\pi(1800)$ as the
appropriate $0^-$ resonance with $m_R=1795\pm 10$ MeV and
$\Gamma=212\pm 37$ MeV \cite{PDG}, we find that ${\cal
B}(D^+\to\rho^+\eta)=3.4\%$ and ${\cal
B}(D^+\to\rho^+\eta')=0.4\%$, which are obviously too large
compared to the current experimental limit: 0.7\% and 0.5\%,
respectively (see Table I). The point is that the $G$ parity of
$\rho\eta$ and $\rho\eta'$ is even, while the $J=0,~I=1$ meson
resonance made from a quark-antiquark pair (i.e. $u\bar d$) has
odd $G$ parity. This is also true for the $W$-annihilation process
$c\bar d\to u\bar d$. As a consequence, the even--$G$ state
$\rho\eta$ or $\rho\eta'$ does not couple to any single meson
resonances, nor to the state produced by the $W$-annihilation
diagram with no gluons emitted by the initial state before
annihilation \cite{Lipkin}.

In the absence of FSI, the branching ratio of $D^+\to \pi^+\eta$
is very small, of order $10^{-4}$, owing to a large cancellation
between external and internal $W$-emission amplitudes, the latter
being enhanced by the fact that $X_s^{(D^+\pi^+,\eta)}\approx
-X_d^{(D^+\pi^+,\eta)}$. Again, owing to the large branching ratio
of $D^+\to K^+\ov K^0$, this mode is essentially induced by FSI
through resonance. Since a nearby $0^+$ resonance $a_0$ in the
charm mass region has not been observed, we employ $m_R=1745$ MeV
and $\Gamma=250$ MeV for calculations.

\subsection{$D_s^+\to (\pi^+,\rho^+)(\eta,\eta')$ decays}
The analysis of resonant coupled-channel effects in $D_s^+\to
K^+\ov K^0,\pi^+\eta_{ns},\pi^+\eta_s$ leads to the replacement of
the $W$-annihilation amplitude by \cite{Zen}:
\be \label{newd}
\d\to \d+\left(\d+{1\over 3}\b\right)(e^{2i\delta_r}-1),
\en
where $\b$ is the internal $W$-emission amplitude for $D_s^+\to
K^+\ov K^0$. As before, neglecting the short-distance
$W$-annihilation, we then have:
\be
A(D_s^+\to\pi^+\eta) &=&
{G_F\over\sqrt{2}}V_{cs}^*V_{ud}\Bigg[a_1X^{(D_s\eta,
\pi^+)}+{1\over 3\sqrt{3}}a_2X^{(D_sK^+,\ov K^0)}  \non \\
&\times& (e^{2i\delta_r}-1)\left(\sqrt{2}\cos\theta-2\sin\theta
\right)\Bigg], \non \\ A(D_s^+\to\pi^+\eta') &=&
{G_F\over\sqrt{2}}V_{cs}^*V_{ud}\Bigg[a_1X^{(D_s
\eta',\pi^+)}+{1\over 3\sqrt{3}}a_2X^{(D_sK^+,\ov K^0)} \non \\
&\times & (e^{2i\delta_r}-1)\left(\sqrt{2}\sin\theta+2\cos\theta
\right)\Bigg].
\en
Unlike $D\to\ov K\etapp,~\pi\etapp$ decays, the resonant FSI here
are induced from internal $W$-emission and hence play a less
significant role. As noted in \cite{CT99}, $D_s^+\to\pi^+\eta$ is
suppressed in the presence of FSI through resonances, whereas
$D_s^+\to\pi^+\eta'$ is enhanced (see Table I). This is ascribed
to the fact that the external $W$-emission amplitudes for
$D_s^+\to\pi^+\eta$ and $\pi^+\eta'$ are opposite in sign due to a
relative sign difference between the form factors $F_0^{D_s\eta}$
and $F_0^{D_s\eta'}$. There are several new measurements of the
$D_s^+$ lifetime \cite{E791}. We use the updated world average
$\tau(D_s^+)=(4.95\pm 0.13)\times 10^{-13}s$ \cite{PDG}.

For reasons to be mentioned below, we shall keep the
$W$-annihilation contribution in $\rho^+\etapp$ decays:
\be
A(D_s^+\to\rho^+\eta) &=&
{G_F\over\sqrt{2}}V_{cs}^*V_{ud}\,a_1\left(X^{
(D_s\eta,\rho^+)}+X^{(D_s,\eta\rho^+)}\right), \non \\
A(D_s^+\to\rho^+\eta') &=&
{G_F\over\sqrt{2}}V_{cs}^*V_{ud}\,a_1\left(X^{
(D_s\eta',\rho^+)}+X^{(D_s,\eta'\rho^+)}\right).
\en
The $W$-annihilation amplitude can be related to the
$D_s^+\to\omega\pi^+$ one via SU(3) symmetry:
\be
A(D_s^+\to\rho^+\eta) &=& {G_F\over\sqrt{2}}V_{cs}^*V_{ud}\,a_1X^{
(D_s\eta,\rho^+)}+\left({1\over\sqrt{3}}\cos\theta-\sqrt{2\over
3}\sin\theta\right)A(D_s^+\to\omega\pi^+),  \non \\
A(D_s^+\to\rho^+\eta') &=&
{G_F\over\sqrt{2}}V_{cs}^*V_{ud}\,a_1X^{
(D_s\eta',\rho^+)}+\left({1\over\sqrt{3}}\sin\theta+\sqrt{2\over
3}\cos\theta\right)A(D_s^+\to\omega\pi^+).
\en
The decay $D_s^+\to\omega\pi^+$, which proceeds through
$W$-annihilation topologies, has been observed recently with the
branching ratio $(0.28\pm 0.11)\%$ \cite{Balest}. Unfortunately,
the phase of this decay relative to $X^{(D_s^+\eta',\rho)}$ is
unknown. In the extreme case that $A(D_s^+\to\omega\pi^+)$ is real
and opposite to $X^{(D_s^+\eta,\rho^+)}$ in sign, then we find
${\cal B}(D_s^+\to\rho^+\eta)=8.3\%$, whereas ${\cal
B}(D_s^+\to\rho^+\eta')=3.6\%$, recalling that the external
$W$-emission amplitudes for $\rho^+\eta$ and $\rho^+\eta'$ are
opposite in sign. By contrast, if the amplitude of $\omega\pi^+$
is purely imaginary, then the branching ratios of $\rho^+\eta$ and
$\rho^+\eta'$ will be 6.4\% and 4.1\%, respectively. Therefore,
even if the former mode is accommodated, the latter is still too
small compared to experiment.

If $\rho^+\eta$ and $\rho^+\eta'$ are dominated by the external
$W$-emission, then it is advantageous to consider the ratios
$R_\etapp\equiv\Gamma(D_s^+\to\rho^+\etapp)/\Gamma(D_s^+\to\etapp
e^+\nu)$. Generalized factorization leads to the
form-factor-independent predictions $R_\eta=2.9$ and
$R_{\eta'}=3.5$, while experimentally $R_\eta=4.4\pm 1.2$ and
$R_{\eta'} =12.0\pm 4.3$ \cite{PDG}. The large discrepancy between
theory and experiment for $R_{\eta'}$ means that there must be an
additional contribution to $\rho^+\eta'$. An enhancement mechanism
has been suggested in \cite{Ball} that a $c\bar s$ pair
annihilates into a $W^+$ and two gluons, then the two gluons will
hadronize  mostly into $\eta'$. If the two gluons couple to
$\etapp$ through the triangle quark loop, they will hadronize into
the flavor-singlet $\eta_0$. Since
\be
\eta_0=\eta'\cos\theta-\eta\sin\theta,
\en
and the $\eta-\eta'$ mixing angle $\theta$ is negative, it is
evident that if $\rho^+\eta'$ is enhanced by this mechanism,
$\rho^+\eta$ will be suppressed due to the destructive
interference between the external $W$-emission and the
gluon-mediated process. Specifically, we find that if the
branching ratio of $\rho^+\eta'$ is increased to 9.5\%, then
${\cal B}(D_s^+\to\rho^+\eta)$ will be decreased to 3.9\%. The
other possibility is that the gluonic component of $\eta'$, which
can be identified with the physical state, e.g. the gluonium,
couples to two gluons directly. From the wave function of the
gluonium \cite{Kou}
\be
{\rm
gluonium}=-\eta\sin\theta\sin\phi-\eta'\cos\theta\sin\phi+g\cos\phi,
\en
where $g$ is a glue rich particle, we see that the gluonium
contribution can enhance both $\rho^+\eta'$ and $\rho^+\eta$,
especially the former; that is, this new mechanism can account for
the unexpectedly large branching ratio of $\rho^+\eta'$ without
suppressing $\rho^+\eta$.

It is clear that a production of the $\eta'$ due to
gluonium-mediated processes can in principle enhance $\rho^+\eta'$
sizeably and $\rho^+\eta$ slightly. Therefore, if the
gluon-mediated process is responsible for the major production of
$\eta'$ in $D_s^+\to\rho^+\eta'$ decay, the two gluons in the
intermediate state must couple to the gluonium rather than to the
$\eta_0$.

Since the additional contribution to $D_s^+\to\rho^+\eta'$ is not
needed to explain the other decays involving $\eta$ and $\eta'$
(see Table I), one may wonder if the new contribution is special
only to the above-mentioned decay. We conjecture that this
mechanism is operative if the naive $W$-annihilation diagram is
prohibited under $G$-parity consideration while allowed when
gluons are emitted from the initial quark. Under this conjecture,
the gluon-mediated processes are important only for the decays
$(D^+,D_s^+)\to\rho^+\eta'$. It is likely that the branching ratio
of $D^+\to\rho^+\eta'$ is enhanced by a factor of 2, namely,
${\cal B} (D^+\to\rho^+\eta')=0.16\%$, which is safely below the
current experimental limit (see Table I).

\section{Theoretical Uncertainties}
The calculation of the effects of resonant FSI suffers from many
theoretical uncertainties. It is useful to explain them below.
\begin{enumerate}
\item
Thus far we have assumed that coupled-channel FSI are dominated by
nearby resonances in the charm mass region; other types of FSI,
e.g. quark exchange, are not considered in the present work.
Resonances with lower masses, e.g. the $1^-$ resonance $K^*(890)$
and the $0^+$ resonant state $K^*(1430)$ have not been included in
our calculations as they are not close to the charm mass region.
However, they can contribute to $W$-exchange or $W$-annihilation
directly via pole diagrams \cite{Gronau,Bedaque}. Recall that the
determination of $a_1$ and $a_2$ from $D\to\ov K\pi$ decays is
usually obtained by neglecting the $W$-exchange contribution. The
inclusion of $K^*(1430)$ resonance will certainly affect the
extraction of $a_1$ and $a_2$ \cite{Gronau}.
\item
For simplicity we have neglected $W$-exchange or $W$-annihilation
contributions in our calculations. However, data analysis based on
the flavor-SU(3) quark-diagram scheme indicates that $W$-exchange
and $W$-annihilation are not negligible \cite{CC87,Rosner}.
\item
Flavor SU(3) symmetry breaking has been introduced to the
couplings $c^{(r)}_{ij}$ in the literature via phase-space
corrections \cite{Buccella,Fajfer}. For example, in \cite{Fajfer}
 phase-space induced SU(3)-symmetry breaking is included in the coupling
constants:
\be
c^{(r)}_{ij}={ \la M_iM_j|H_{\rm eff}|r\ra \sqrt{p_{ij}}\over
\sqrt{\sum_{ij}\la M_iM_j|H_{\rm eff}|r\ra^2\, p_{ij}} },
\en
where $p_{ij}$ is the momentum of the final particles in the $D$
rest frame. In our work, we assume SU(3) flavor symmetry for
quark-diagram  amplitudes and consider its breaking only at the
decay rate level; it seems to us that it is not appropriate to
have the phase-space correction in coupling constants.
\item
Recently it was found in \cite{Kroll2} that phenomenologically the
$\eta-\eta'$ mixing angle is given by $\theta=-15.4^\circ$, which
is somewhat smaller than the mixing angle $-22^\circ$ employed in
the present paper. However, we found empirically that the latter
yields a better agreement between theory and experiment than the
former. For example, in the absence of resonant FSI, or
equivalently $\theta=-19.5^\circ$, ${\cal B}(D^0\to\ov
K^0\eta)=0.54\%$. The branching ratio is increased to 0.62\% when
$\theta=-22^\circ$ and decreased to 0.35\% at
$\theta=-15.4^\circ$, recalling that experimentally ${\cal
B}(D^0\to\ov K^0\eta)=(0.70\pm 0.10)\%$ \cite{PDG}. Of course, we
cannot conclude that the magnitude of the mixing angle should be
larger than $20^\circ$ in view of many simplified assumptions we
have made.
\item
Resonance-induced FSI are mainly governed by the width and the
mass of nearby resonances, which are unfortunately not well
determined. For example, a reanalysis in a $K$-matrix formalism
\cite{Anisovich} quotes $m_R=1820\pm 40$ MeV and $\Gamma=250\pm
50$ MeV for the $0^+$ resonance $K^*(1950)$. We then obtain ${\cal
B}(D^0\to\ov K^0\eta)=0.65\%$ and ${\cal B}(D^0\to\ov
K^0\eta')=3.44\%$, to be compared with 0.62\% and 2.57\%,
respectively for $m_R=1945$ MeV and $\Gamma=210$ MeV. Hence, the
prediction of ${\cal B}(D^0\to\ov K^0\eta)$ is significantly
affected by the uncertainties in $m_R$ and $\Gamma$ of the
resonance.

\end{enumerate}

\section{Conclusions}
For hadronic decay modes $D^0\to(\ov K^0,\ov K^{*0})(\eta,\eta')$
and $(D^+,D_s^+)\to(\pi^+,\rho^+)(\eta,\eta')$ which have only one
single isospin component, resonance-induced final-state
interactions that mimic the $W$-exchange or the $W$-annihilation
topology can play an essential role. In particular, the decays
$D^0\to\ov K^0\eta'$ and $D^+\to\pi^+\eta$, which are largely
suppressed in the absence of FSI, are enhanced dramatically by
resonant FSI. It is stressed that resonant FSI are negligible for
$\rho^+(\eta,\eta')$ final states because of the mismatch of the
$G$ parity of $\rho^+(\eta,\eta')$ and the $J=0$, $I=1$ meson
resonance.

We have utilized the mode $D^0\to\ov K^0\pi^0$ as an illustration
to demonstrate that resonance-induced FSI, which are crucial for
some two-body decays involving one single isospin component, e.g.
the final state containing an $\eta$ and $\eta'$, play only a
minor role compared to isospin FSI for decay modes involving
several different isospin components.

It is difficult to understand the observed large decay rates of
$D_s^+\to \rho^+\eta'$. We argue that a possible  gluon-mediated
process in which the two gluons couple directly to the gluonic
component of the $\eta'$, e.g. the gluonium, rather than to the
flavor-singlet $\eta_0$, can enhance both decays
$D_s^+\to\rho^+\eta'$ and $D_s^+\to\rho^+\eta$, especially the
former. Since this additional contribution is not needed to
explain the other decays involving the $\eta$ and $\eta'$, we
conjecture that this mechanism is operative if the naive
$W$-annihilation is prohibited under $G$-parity consideration
while allowed when gluons are emitted from the initial quark.

\vskip 1.5 cm \acknowledgments  This work was supported in part by
the National Science Council of ROC under Contract No.
NSC89-2112-M-001-016.

%%%%% References %%%%%%%%%%%%%%%%%%%%%%%%%%%%%%%%%%%%%%%%%%%%%%%%%%%%%%%%%%%%%

\newcommand{\bi}{\bibitem}

\newpage
\vskip 0.1cm
\begin{table}[ht]
{\small Table I. Branching ratios (in units of \%) of the charmed
meson decays to an $\eta$ or $\eta'$.}
\begin{center}
\begin{tabular}{ l c c c c }
 & \multicolumn{2}{c}{This work} & \\\cline{2-3}
\raisebox{1.5ex}[0cm][0cm]{Decay} &  without FSI & with resonant
FSI & \raisebox{1.5ex}[0cm][0cm] {Buccella {\it et al.}
\cite{Buccella}} & \raisebox{1.5ex}[0cm] [0cm]{Expt. \cite{PDG} }
\\ \hline $D^0\to\ov K^0\eta$ & 0.54 & 0.62 & 0.84 & $0.70\pm
0.10$ \\ $D^0\to\ov K^0\eta'$ & 0.10 & 2.57 & 1.56 & $1.71\pm
0.26$\\ $D^0\to\ov K^{*0}\eta$ & 0.69 & 0.81 & 0.37 & $1.9\pm 0.5$
\\ $D^0\to\ov K^{*0}\eta'$ & 0.004 & 0.05 & 0.004 & $<0.10$   \\
\hline $D^+\to\pi^+\eta$ & 0.02 & 0.31 & 0.34 & $0.30\pm 0.06$
\\ $D^+\to\pi^+\eta'$  & 0.29 & 0.72 & 0.73 & $0.50\pm 0.10$  \\
$D^+\to\rho^+\eta$ & 0.19 & 0.19 & 0.013 & $<0.7$  \\
$D^+\to\rho^+\eta'$ & 0.08 & 0.08 & 0.12 & $<0.5$  \\ \hline
$D_s^+\to\pi^+\eta$ & 2.57 & 1.95 & 1.30 & $1.7\pm 0.5$ \\
$D_s^+\to\pi^+\eta'$ & 3.50 & 4.28 & 5.71 & $3.9\pm 1.0$ \\
$D_s^+\to\rho^+\eta$ & 6.27 & 6.27$^*$ & 7.94 & $10.8\pm 3.1$ \\
$D_s^+\to\rho^+\eta'$ & 4.09 & 4.09$^*$ & 2.55 & $10.1\pm 2.8$ \\
\end{tabular}
\noindent $^*$ The presence of $W$-annihilation contributions
inferred from $D_s^+\to\omega\pi^+$ will affect the branching
ratios of $D_s^+\to\rho^+\etapp$; see the text.
\end{center}
\end{table}

\end{document}